\documentclass[prl,aps,showpacs,twocolumn]{revtex4}

\pdfoutput=0

\usepackage{soul,xcolor}
\usepackage{amsmath}
\usepackage{amsfonts}
\usepackage{graphicx,subfigure,color}
\usepackage{bm}
\usepackage{color}

\def\be{\begin{equation}}
\def\ee{\end{equation}}

\begin{document}

\setstcolor{blue}

\title{
Spin Drag of a Fermi Gas in a Harmonic Trap}
\author{O. Goulko$^1$, F. Chevy$^2$, C. Lobo$^3$}
\affiliation{
$^1$Physics Department, Arnold Sommerfeld Center for Theoretical Physics, and Center for NanoScience,
Ludwig-Maximilians-Universit\"at, Theresienstra\ss e 37, 80333 Munich, Germany\\
$^2$Laboratoire Kastler Brossel, CNRS, UPMC, \'Ecole Normale Sup\'erieure, 24 rue Lhomond, 75231 Paris, France\\
$^3$School of Mathematics, University of Southampton, Highfield, Southampton, SO17 1BJ, United Kingdom
}
\date{\today}

\begin{abstract}

Using a Boltzmann equation approach, we analyze how the spin drag of a trapped interacting fermionic mixture is influenced by the non-homogeneity of the system in a classical regime where the temperature is much larger than the Fermi temperature. We show that for very elongated geometries, the spin damping rate can be related to the spin conductance of an infinitely long cylinder. We characterize analytically the spin conductance both in the hydrodynamic and collisionless limits and discuss the influence of the velocity profile. Our results are in good agreement with recent experiments and provide a quantitative benchmark for further studies of spin drag in ultracold gases.

\end{abstract}

\maketitle

In the recent years, ultracold atoms have become a unique testing ground for quantum many-body physics. Their study has favored the emergence of novel experimental and theoretical techniques which have led to an accurate quantitative understanding of the thermodynamic properties of strongly correlated dilute gases at equilibrium \cite{zwerger2012BCSBEC}. An important effort is now devoted to the exploration of the out-of-equilibrium behavior of these systems, and in particular to the determination of their transport properties. For instance, recent experiments have probed the transport of an ultracold sample through a mesoscopic channel \cite{brantut2012conduction}, and time of flight expansions have been used to measure the gas viscosity in the strongly correlated regime \cite{cao2010observation} where it is predicted to be close to the universal limit conjectured by string theory \cite{kovtun2005viscosity}.

In this Letter we focus on spin transport properties of a Fermi gas which have now received considerable attention in the cold atom community \cite{liao2011metastability,bruun2012shear,enss2012quantum,wong2012spin,heiselberg2012inhomogeneous,kittinaradorn2012critical,kim2012heat} after previously being studied in the context of liquid $^3$He \cite{Meyerovich}, ferromagnetic metals \cite{Mineev} and spintronic materials \cite{wolf2001spintronics}. Recent measurements of the spin drag coefficient \cite{sommer2011universal,summer2011spin} have shown that the most challenging aspect of these studies is how to extract the homogeneous gas properties from measurements performed in harmonic traps. The trapping potential creates a density inhomogeneity which can significantly alter the transport behaviour of the gas, because the local mean free path can vary strongly from point to point in the trap leading to a coexistence of regions, from hydrodynamic near the cloud center to collisionless at the edge \cite{bruun2011spinb}. For the same reason, the  velocity during the relaxation to equilibrium is not constant as a function of radius and it is essential that it be accurately known in order to find the correct values of transport coefficients. Previous theoretical attempts to cope with these problems have included making unverified assumptions about the velocity profile of the gas \cite{vichi1999collective,chiacchiera2010dipole,bruun2011spin} or treating the problem in the hydrodynamic approximation with spatially varying spin diffusivity \cite{bruun2011spinb}. In this last work, no quantitative conclusion could be obtained due to the importance of the collisionless regions of the cloud.

Here we present a systematic study of the spin transport in an elongated harmonic trap based on the Boltzmann equation using a combination of analytical and numerical methods in the dilute limit and for small phase-space density. In this regime we are able to analyze the behavior of the trapped gas, allowing us to deal {\it ab initio} with the spatial density changes without any uncontrolled approximations. In particular we are able to make definite predictions for the spin drag coefficient and the transverse velocity profile in both the collisionless and hydrodynamic regimes. Due to the fact that the trapping is much weaker in the axial than the radial direction, we can use the local density approximation to relate the local spin conductance in each slice perpendicular to the axis of the trap to the spin conductance in an infinite trap with the same central density.

Consider an ensemble of spin $1/2$ fermions of mass $m$ confined in a very elongated harmonic trap with axial frequency $\omega_z$ and transverse frequency $\omega_x=\omega_y\equiv\omega_\perp\gg\omega_z$. Each atom has $s=\pm$ spin  with equal numbers of atoms in each spin state. In the initial thermal equilibrium state the two spin species are separated from each other by an average distance of $\pm z_0$ along the symmetry axis of the trap as in \cite{sommer2011universal}. Then we let the system relax towards equilibrium and, as observed experimentally, the relaxation of the motion of the centers of mass of the two clouds occurs at a rate $\propto \omega_z^2/\gamma_{\rm coll}$, where  $\gamma_{\rm coll}$ is the collision rate \cite{sommer2011universal}. In the very elongated limit $\omega_z\ll \omega_\perp,\gamma_{\rm coll}$ the momentum and spatial transverse degrees of freedom are therefore always thermalized and we can assume that the phase space density of the spin species $s=\pm$ is given by the ansatz
\be
f_s(\bm r,\bm p,t)=f_0(\bm r,\bm p)(1+s\alpha(z,t)),
\label{Eq1b}
\ee
where $f_0$ is the equilibrium phase-space density. As long as interparticle correlations are weak, the single particle phase space density encapsulates all the statistical information on the system. In the rest of the Letter we will restrict ourselves to such a regime. Since the experiment \cite{sommer2011universal} was performed at unitarity, this condition is achieved when the temperature is much larger than the Fermi temperature. As a consequence, we can also neglect Pauli blocking during collisions.

Let $\bar n_s(z,t)=\int d^2\bm\rho d^3\bm p f_s(\bm r,\bm p,t)=\bar n_0(z) (1+s\alpha(z,t))$ be the 1D-density along the axis of the trap, where $\bm\rho=(x,y)$. Due to particle number conservation we have
\be
\partial_t\bar n_s+\partial_z\Phi_s=0,
\label{Eq2b}
\ee
where $\Phi_s=\int d^2\bm\rho d^3\bm p f_s(\bm r,\bm p)v_z$ (with $v_z=p_z/m$, the axial velocity) is the particle flux of spin $s$ in the $z$ direction. If the trap is very elongated we can define a length scale $\ell$ much smaller than the axial size of the cloud, but much larger than its transverse radius, the interparticle distance or the collisional mean-free path so that for distances smaller than $\ell$ along the $z$ axis,  the physics can be viewed as being equivalent to that of an infinitely elongated trap ($\omega_z=0$) with the same central density. In this setup the two spin species are pulled apart by a force $\bm F_s=-\bm\nabla V-(\bm\nabla P_s)/n_s$ where $V$ is the spin independent trapping potential, $P_s$ is the pressure of the spin species $s$ and $n_s(\bm r,t)=\int d^3\bm p f_s(\bm r,\bm p,t)$ is the associated density. We consider here a classical ideal gas, for which $P_s=nsk_BT$. Using the ansatz (\ref{Eq1b}), we see that the force field is uniform and is given by $\bm F_s=-sk_BT\partial_z \alpha\bm e_z \equiv F_s \bm e_z$, where $\bm e_z$ is the unit vector along $z$, since $\partial_z\alpha$ can be considered constant to leading order on the length scale $\ell$.

\begin{figure}
\centerline{\includegraphics[width=\columnwidth]{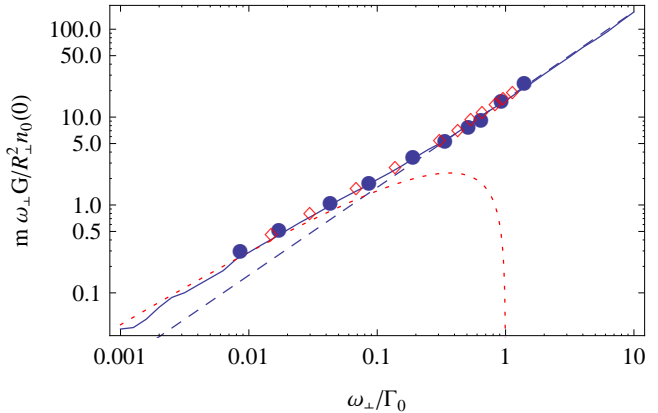}}
\caption{Color online.
Spin conductance $G$ vs. $\omega_\perp/\Gamma_0$. The blue dashed line corresponds to the collisionless limit $G\simeq 15.87 n_0/\Gamma_0$ for a Maxwellian gas while the red-dotted line is the hydrodynamic prediction $G\simeq (2\pi n_0/\Gamma_0) \ln\Gamma_0$. The solid line is the semi-analytical prediction for the Maxwellian gas (see Supplemental Material). The blue dots are the results of the molecular dynamics simulation for a constant cross-section. The red open diamonds correspond to a momentum dependent cross-section with $k_{\rm th}a=2$, where $k_{\rm th}=\sqrt{mk_BT/\hbar^2}$ is the thermal wavevector.
}
\label{Fig1}
\end{figure}

In the regime of linear response, the particle flux is proportional to the drag force and we can write $\Phi_s=GF_s$, where $G$ is the ``spin conductance" that {\em a priori} depends on the 1D-density of the cloud. Inserting this law in Eq. (\ref{Eq2b}) and substituting $\alpha(z,t)=e^{-\gamma t}\alpha_0(z)$, which corresponds to the exponential decay of the perturbation close to equilibrium, we see that $\alpha_0$ is solution of
\be
\gamma\bar n_0(z) \alpha_0(z)+k_B T\partial_{z}\left(G(\bar n_0(z))\partial_{z}\alpha_0(z)\right)=0.
\label{Eq:5}
\ee
The exponential coefficient $\gamma$ defines the decay time close to equilibrium and thus the spin drag.  This equation can be derived more rigorously from a systematic expansion of Boltzmann's equation (see Supplemental Material) and is equivalent to the Smoluchowski equation derived in \cite{bruun2011spinb} if one takes for the spin diffusion coefficient $D=k_{\rm B} T G/\bar n$. Eq. (\ref{Eq:5}) is supplemented by the condition  $\Phi_s(\pm\infty)=0$ imposed by particle number conservation. Since, as we will show below, the spin conductance is a (non-zero) constant in the dilute limit, this constraint yields the boundary condition $\partial_z \alpha_0=0$ at $z=\pm\infty$.
\begin{figure}
\centerline{\includegraphics[width=\columnwidth]{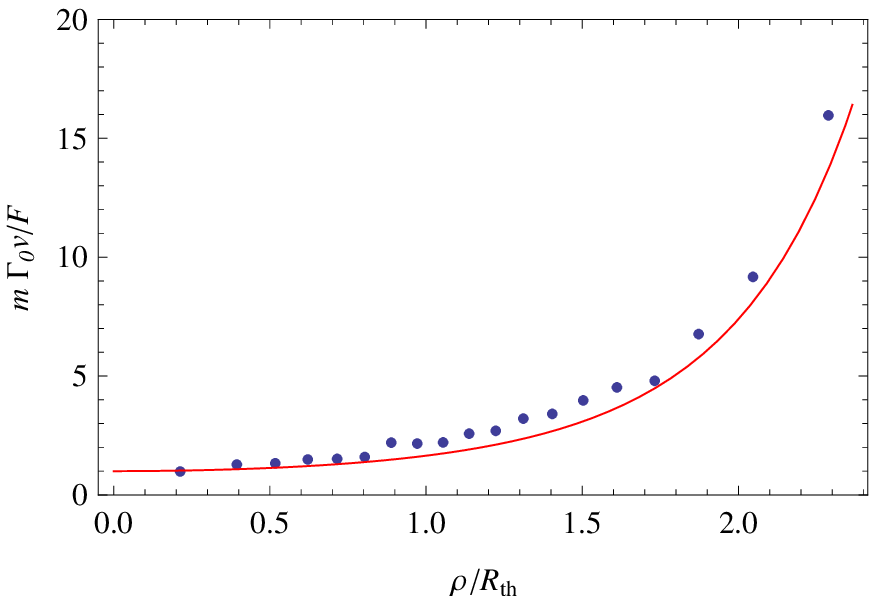}}
\centerline{\includegraphics[width=\columnwidth]{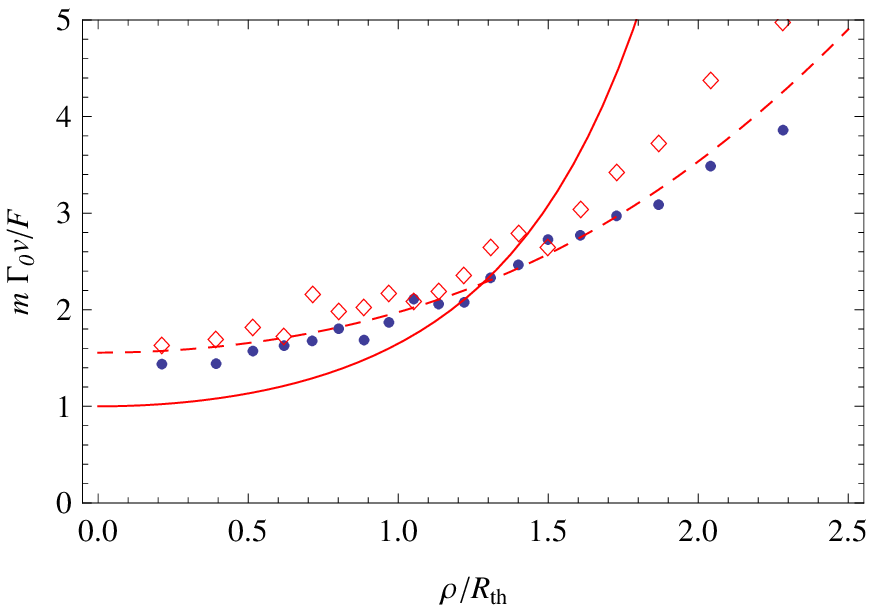}}
\caption{Color online. Transverse velocity field $v_s(\rho)$ in units of $m\Gamma_0/F$ in the infinitely elongated trap. Top: hydrodynamic regime, $\Gamma_0/\omega_\perp\simeq 100$; Bottom: collisionless regime $\Gamma_0/\omega_\perp\simeq 2$.
The blue dots are simulation results for the constant scattering cross-section and the red solid line is the prediction $v_s(\rho)=F_s/m\Gamma(n_0(\rho))$ for the hydrodynamic regime. On the lower graph, the empty red diamonds are simulation results for the momentum dependent cross-section at $k_{\rm th}a=2$ and the red-dashed line represents the velocity field of a Maxwellian gas in the collisionless limit (see Supplemental Material). }
\label{Fig3}
\end{figure}

Before solving this equation to find $\gamma$, we need to know the expression of the spin conductance $G$. We first consider the simpler case of a uniform gas of density $n_+=n_-=n_0=$~const. Using the method of moments \cite{vichi1999collective}, the velocity is solution of the equation $\partial_t v_s+\Gamma(n_0)v_s=F_s/m$, where the spin damping rate $\Gamma$ is given by
\be
\Gamma(n_0)=\frac{1}{n_0}\int  d^3\bm p f_{0}^{\rm (H)} (\bm p) p_z C[p_z],
\label{Eq3b}
\ee
where $f_{0}^{\rm (H)}(\bm p)=n_0 e^{-p^2/2mk_B T}/(2\pi m k_B T)^{3/2}$ is the Gaussian phase-space density of a homogeneous gas and $C[\alpha]$ is the linearized collisional operator defined by
\be
C[\alpha](\bm p_1)=\int d^3\bm p_2 f_0^{\rm (H)}(\bm p_2)v_{\rm rel}\sigma(v_{\rm rel})\left(\alpha_2-\alpha_1\right),
\ee
where $v_{\rm rel}=|\bm p_2-\bm p_1|/m$, $\sigma$ is the s-wave scattering cross-section and $\alpha_i$ stands for $\alpha(\bm p_i)$ \footnote{Strictly speaking the expression for $\Gamma$ in Eq. (\ref{Eq3b}) was obtained using an uncontrolled ansatz for the phase-space density. Using a molecular dynamics simulation, we checked that this ansatz does indeed yield very accurate results for the homogeneous gas.}. Generally speaking, $\Gamma$ is proportional to the collision rate, with a numerical prefactor depending on the actual form of the scattering cross-section.
In the homogeneous case the stationary velocity is simply given by $v_s=F_s/m\Gamma(n_0)$. In a trap, the density profile is inhomogeneous, which leads to a shear of the velocity field and a competition between viscosity and spin drag. Let $R_{\rm th}=\sqrt{k_B T/m\omega_\perp^2}$ be the transverse size of the cloud and  $\nu$ its kinematic viscosity. Viscosity can be neglected as long as the viscous damping rate $\nu/R_{\rm th}^2$ is smaller than $\Gamma(n_0)$. Since  viscosity scales like $v_{\rm th}^2/\gamma_{\rm coll}$, with the thermal velocity $v_{\rm th}=\sqrt{k_B T/m}$, this condition is fulfilled as long as $\Gamma\propto\gamma_{\rm coll}\gg \omega_\perp$, in other words when the cloud is hydrodynamic in the transverse direction. In this regime we can therefore neglect viscous stress and the local velocity $v_s(\bm \rho)=\int d^3\bm p f_s(\bm\rho,\bm p)v_z/n_s(\bm\rho)$ is simply given by $v_s(\bm \rho)=F_s/m\Gamma(n_0(\bm\rho))$, where $n_0(\rho)=n_0(0)\exp(-\rho^2/2R_{\rm th}^2)$ is the local equilibrium density of the cloud.

This scaling for the velocity field is however too simple. Indeed, we have $\Phi_s=\int d^2\bm\rho n_0(\bm \rho)v_s(\bm\rho)\propto \int d^2\bm\rho n_0/\Gamma(n_0(\bm\rho))$, and since $\Gamma\propto n_0$, the integral is divergent. This pathology is cured by noting that the hydrodynamic assumption is not valid in the wings of the distribution where the density, and therefore the collision rate, vanish. The breakdown of the hydrodynamic approximation occurs when $\Gamma (n_0(\rho))\lesssim \omega_\perp$, i.e.\ when $\rho\gtrsim \rho_{\rm max}=R_{\rm th}\sqrt{2\ln (\Gamma_0/\omega_\perp)}$, with $\Gamma_0=\Gamma(n_0(0))$ the local spin damping at the trap center. Considering $\rho_{\rm max}$ as a cut-off in the integral for $G$ we see that $G\simeq 2\pi \rho_{\rm max}^2 n_0(0)/m\Gamma_0\propto \ln(\Gamma_0)/\Gamma_0$.

In the opposite regime, when the gas is collisionless in the transverse direction, we expect viscous effects to flatten the velocity profile. Assuming a perfectly flat velocity field, then $v_s\propto F_s/m\Gamma_0$ and thus $G=\Phi_s/F_s \propto \bar n_0/m\Gamma_0$.

To make this scaling argument more quantitative, we calculate $G$ for different physical situations. First we calculate it numerically using the Boltzmann equation simulation described in \cite{goulko2011collision,goulko2012boltzmann}. We initialise the axially homogeneous system at thermal equilibrium and then switch on the constant pulling force at $t=0$. We observe that in a few collision times, the total spin current of the cloud defined by $\Phi_s(t)=\langle v_z\rangle_s=\int d^3\bm r d^3\bm p f_s(\bm r,\bm p,t) v_z$ converges to a constant asymptotic value from which we extract the spin conductance $G(\bar n_0)$. Figure~\ref{Fig1} shows our results for the spin conductance for a constant cross section $\sigma=4\pi a^2$ and a momentum-dependent cross section $\sigma=4\pi a^2/(1+p_{\rm rel}^2 a^2/4)$ near the unitary limit \footnote{For practical reasons, we limited our study of the strongly interacting regime to $k_{\rm th}a=2$. For this value, the difference with the unitary gas prediction for the value of $\Gamma$ is only 10\%.}. When $G/n_0(0)$ is plotted versus $\Gamma_0=\Gamma(n_0(0))$ the data points overlap, showing that the drag coefficient depends only weakly on the actual momentum dependence of the scattering cross-section. To interpolate between the constant and the unitary cross section we also study the Maxwellian cross section $\sigma\propto 1/p$ for which we could find a semi-analytical expression of the spin conductance (see Supplemental Material).

Using these approaches, we find the following asymptotic behaviors. As expected, in the (transverse) collisionless limit $\Gamma_0\ll \omega_\perp$ the spin drag coefficient scales like $G\simeq k n_0(0)R_{\rm th}^2/m\Gamma_0$, where $k\simeq 16$ is a numerical coefficient, the value of which depends on the momentum dependence of the scattering cross-section (see Table \ref{Table1}). In the case of a Maxwellian gas, we find that $k=15.87$ (see Supplemental Material). For more general cases,  a variational lower bound based on the exact Maxwellian solution yields an estimate very close to the numerical result obtained from the molecular dynamics simulation. In the opposite (hydrodynamic) limit $\Gamma_0\rightarrow\infty$, we recover the expected behavior $G\simeq 2\pi n_0(0)R_{\rm th}^2 \ln(\Gamma_0/\omega_\perp)/m\Gamma_0$.
\begin{table}
\begin{tabular}{cccc}
$n$&$0$&$-1$&$-2$\\
\hline
\hline
Variational lower bound&14.5&15.87&17\\
Molecular dynamics&15.4& - &18.9\\
\hline
\end{tabular}
\caption{
Values of $k$ for a scattering cross-section $\sigma (p)\propto p^n$ for a constant cross-section ($n=0$), a Maxwellian gas ($n=-1$), and a unitary gas $(n=-2)$. For the Maxwellian gas, the lower bound is actually the exact result.}
\label{Table1}
\end{table}

We also calculate the transverse velocity profile $v_s(\bm\rho)$ and confirm that it obeys the expected behavior, see Fig.~\ref{Fig3}.
For $\Gamma_0/\omega_\perp\gg 1$, we recover the viscousless prediction $v_s\propto 1/\Gamma (n_0(\bm\rho))$ while for $\Gamma_0\lesssim\omega_\perp$ we obtain a flatter velocity profile as a result of the transverse shearing. We see that in both regimes the velocity profile is not flat, and this explains the discrepancy between experiment and previous theoretical models based on uniform velocities.

\begin{figure}
\centerline{\includegraphics[width=\columnwidth]{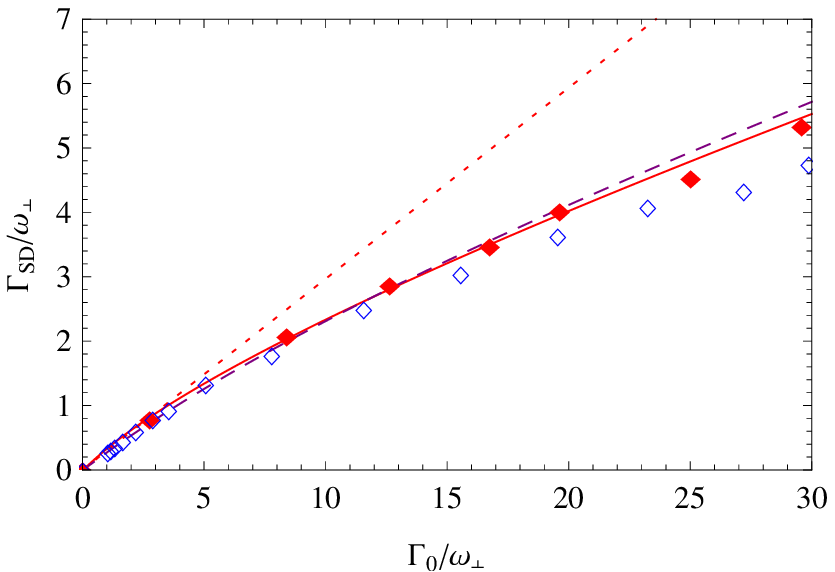}}
\centerline{\includegraphics[width=\columnwidth]{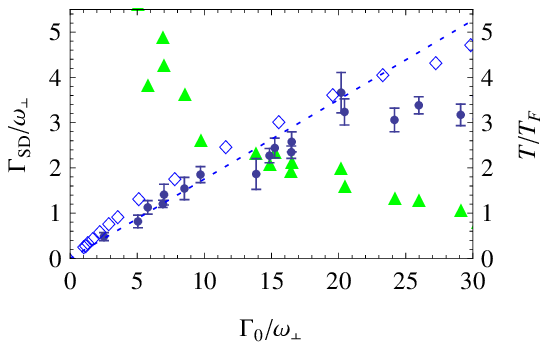}}
\caption{(Color online). Spin-drag coefficient $\Gamma_{\rm SD}$ in a harmonic trap. Top: theoretical predictions. For a constant cross-section the red solid diamonds are the results of the molecular dynamics simulation and the red solid line is the numerical resolution of the eigenequation (\ref{Eq:5}). For a momentum dependent cross-section close to the unitary limit ($k_{\rm th}a =2$, with $k_{\rm th}=\sqrt{k_B T/\hbar^2}$ the thermal wave-vector), the blue open diamonds correspond to the molecular dynamics simulation while the the blue dashed line is the solution of Eq. (\ref{Eq:5}). The red dotted line is the collisionless prediction $\Gamma_{\rm SD}=\Gamma_0/4.03$ for a unitary gas.
Bottom: comparison with the experimental results of Ref. \cite{sommer2011universal} at unitarity (blue dots). The green triangles represent the associated values of $T/T_{\rm F}$. As above, the blue open diamonds correspond to the molecular dynamic simulation at $k_{\rm th}a =2$. Blue-dotted line: experimental fit $\Gamma_{\rm SD}\simeq \Gamma_0/5.7$. For the experimental data, $\Gamma_0$ is calculated using the theoretical value Eq. (\ref{Eq3b}) for the unitary gas.}
\label{Fig2}
\end{figure}

Let us now return to the case of a three-dimensional trap and to the determination of the spin damping rate $\gamma$. According to Eq. (\ref{Eq:5}) $\gamma$ appears as an eigenvalue of the operator $\widehat S=k_B T\bar n_0^{-1}\partial_{z}\left(G(z)\partial_{z}\cdot\right)$. This operator is hermitian on the Hilbert space of functions having a finite limit and zero derivative at $z=\pm\infty$ and since at long times the decay is dominated by the slowest mode, we focus on its smallest eigenvalue. We first consider the collisionless limit. In this regime, $G\propto n_0/\Gamma_0$ is position independent and can be considered as constant. Using the shooting method \cite{press2007numerical} we then obtain
\be
\gamma\simeq 1.342\frac{k \omega_z^2}{2\pi \Gamma_0}.
\label{Eq7}
\ee
where the value of $k$ is given in Table \ref{Table1}. For arbitrary values of $\Gamma_0/\omega_\perp$, we solve Eq. (\ref{Eq:5}) using for $G$ a Pad\'e interpolation of the simulation results presented in Fig. \ref{Fig1} (see Supplemental Material). Following \cite{sommer2011universal}, we take $\Gamma_{\rm SD}=\omega_z^2/\gamma$ and in Fig.~\ref{Fig2}a we plot $\Gamma_{\rm SD}/\omega_\perp$ vs. $\Gamma_0/\omega_\perp$. We compare our model to the experimental results  of Ref. \cite{sommer2011universal} and to a direct molecular dynamics simulation of the Boltzmann equation \cite{goulko2011collision}. In this simulation, the atoms are prepared in a harmonic trap of axial frequency $\omega_\perp=8\omega_z$. We displace their centers of mass by a distance $\pm z_0$, where $z_0$ is much smaller than the axial size of the cloud, and we fit the relative displacement vs time to an exponential from which we extract $\Gamma_{\rm SD}$. The results of these simulations are displayed in Fig. \ref{Fig2}a where they are compared to the solutions of Eq. (\ref{Eq:5}). We observe that the two approaches coincide both for the constant and momentum-dependent cross-sections \footnote{In the case of the momentum dependent cross-section, we observe a $\simeq$10~\% deviation at large collision rate that we interpret as resulting from a systematic error of the same order of magnitude introduced by the Pad\'e approximation of the spin conductance.}.

As observed in Fig. \ref{Fig2}b, theory and experiment agree remarkably as long as $T/T_{\rm F}\gtrsim 2$. Beyond that limit, we enter the quantum degenerate regime where the Boltzmann equation is no longer valid and, as expected, we observe that experiment and theory deviate from each other. In the high-temperature, collisionless limit, we find for the ``unitary" value $k=18.9$, $\Gamma_{\rm SD}\simeq\Gamma_0/4.03$. This result differs from the high-temperature value $\Gamma_{\rm SD}\simeq\Gamma_0/5.7$ found in \cite{sommer2011universal}. We interpret this discrepancy by noting that the theoretical asymptotic behavior Eq. (\ref{Eq7}) is valid for $\Gamma_0/\omega_\perp \lesssim 5$, while the experimental value was obtained by linear-fitting the points with $\Gamma_0/\omega_\perp\lesssim 15$, i.e.   in a regime where the gas was likely less collisionless. Fitting our data on the same scale using a linear law would indeed give $\Gamma_{\rm SD}\simeq \Gamma_0/5.0$. We also note that our scaling $\Gamma_{\rm SD}=\omega_\perp f(\Gamma_0/\omega_\perp)$ contradicts the scaling $\hbar\Gamma_{\rm SD}=E_Fg(T/T_F)$, where $E_F$ and $T_F$ are the Fermi energy and temperature, used in Ref. \cite{sommer2011universal} to analyze the experimental data. The two scalings agree only in the collisionless limit where $f$ is linear, hence outside of the region explored by experiments.

In summary, we have studied the classical dynamics of spin transport in a trap using the Boltzmann equation approach. By taking into account {\em ab initio} the trap inhomogeneity we are able to reproduce the experimental results without uncontrolled approximations and obtain several robust results which allow for a more rigorous extraction of transport coefficients from measurements in trapped cold gases. We highlight the competition between viscosity and spin drag in the shape of the velocity profile which is a crucial ingredient in the understanding of transport properties in a trap. We also demonstrate the breakdown of the universal scaling used to interpret the data of Ref. \cite{sommer2011universal}  in the experimentally relevant range of parameters. In the future we anticipate extending this approach to lower temperatures where many-body interactions and Pauli blocking play a significant role.

We thank M.~Zwierlein and A.~Sommer for fruitful discussions and for providing us with the experimental data of Fig.~\ref{Fig2}. OG acknowledges support from the Excellence Cluster ``Nanosystems Initiative Munich (NIM)". FC acknowledges support from ERC (Advanced grant Ferlodim and Starting grant Thermodynamix), R\'egion Ile de France (IFRAF) and Institut Universitaire de France. CL acknowledges support from EPSRC through grant EP/I018514.

\bibliography{bibliographie}

\newpage

\section{Supplemental Material to Spin Drag of a Fermi Gas in a Harmonic Trap}



\section{Derivation of the transport equation from Boltzmann's equation}

We consider an ensemble of spin 1/2 fermions of mass $m$. In the dilute limit, the statistical properties of the system are fully captured by the single-particle phase-space densities $f_s (\bm r,\bm p,t)$  of the spin species $s=\pm$. In the presence of a trapping potential $V$, the evolution of $f_s$ is given by Boltzmann's equation

\be
\partial_t f_s+\frac{\bm p}{m}\cdot\partial_{\bm r}f_s+\bm F\cdot\partial_{\bm p}f_s=I_{\rm coll}[f_s,f_{-s}],
\label{Eq:1}
\ee
where $\bm F=-\partial_{\bm r}V$ is the trapping force and $I_{\rm coll}$ is the collision operator. For low-temperature fermions, collisions between same-spin particles are suppressed and at low phase-space densities, the collision operator is given by
\be
\begin{split}
&I_{\rm coll}[f_s,f_{-s}](\bm r,\bm p_1)=\\
&\int d^3\bm p_2 d^2\bm\Omega'v_{\rm rel}\frac{d\sigma}{d\Omega'}\left(f_{s,3}f_{-s,4}-f_{s,1}f_{-s,2}\right),
\end{split}
\ee
where $\bm p_1$ and $\bm p_2$ ($\bm p_3$ and $\bm p_4$) are ingoing (outgoing) momenta satisfying energy and momentum conservation, $v_{\rm rel}=|\bm p_2-\bm p_1|/m$ is the relative velocity, $d\sigma/d\Omega'$ is the differential cross-section towards the solid angle $\bm\Omega'$ and  $f_{s,i}$ stands for  $f_s(\bm r,\bm p_i)$.

When the populations of the two spin states are equal, the equilibrium solution of Eq.~(\ref{Eq:1}) is given by the Maxwell-Boltzmann distribution $f_+=f_-=f_0 ={\cal A}\exp\left[-\beta(p^2/2m+V)\right]$, where $\beta=1/k_B T$ and ${\cal A}$ is an integration constant such that $\int d^3\bm rd^3\bm p f_0$ is the population of one spin state. We consider a spin perturbation of the form $f_s(\bm r,\bm p,t)=f_0(\bm r,\bm p)\left(1+s \alpha(\bm r,\bm p,t)\right)$. Assuming the perturbation is small enough, we can expand Boltzmann's equation in $\alpha$ and to first order we obtain
\be
\partial_t \alpha+\frac{\bm p}{m}\cdot\partial_{\bm r}\alpha+\bm F\cdot\partial_{\bm p}\alpha=-C[\alpha],
\label{Eq:2}
\ee
where, for s-wave collisions, the linearized collisional operator $C$ is given by
\be
C[\alpha](\bm r,\bm p_1)=\int d^3\bm p_2 f_0(\bm r,\bm p_2)v_{\rm rel}\sigma(v_{\rm rel})\left(\alpha_1-\alpha_2\right),
\ee
and as above $\alpha_i=\alpha(\bm r,\bm p_i)$.
In experiments, the trap can be  described by a cylindrically-symmetric harmonic potential with frequency $\omega_z$ along the symmetry axis $z$ and $\omega_\perp$ in the transverse $(x,y)$ plane. In the rest of this Supplemental Material, we work in a unit system where $m=k_BT=\omega_\perp=1$ and we then write $V(x,y,z)=(\omega_z^2z^2+\rho^2)/2$, with $\bm\rho=(x,y)$.

We look for exponentially decaying solutions corresponding to small deviations from equilibrium, and therefore take $\alpha(\bm r,\bm p,t)=e^{-\gamma t}\widetilde\alpha (\bm r,\bm p)$. Eq.~(\ref{Eq:2}) then becomes
\be
\left[-\gamma+p_z\partial_z-\omega_z^2z\partial_{p_z}\right]\widetilde\alpha =-\left[\bm\Pi\cdot\partial_{\bm\rho}-\bm\rho\cdot\partial_{\bm\Pi}+C\right]\widetilde\alpha,
\label{Eq:3}
\ee
where $\bm\Pi=(p_x,p_y)$ is the projection of the momentum in the $(x,y)$ plane. We note that in the rhs of Eq. (\ref{Eq:3}) the only $z$-dependence is in the linearized collisional operator $C$, from $f_0\propto \exp(-\omega_z^2 z^2/2)$. Let $C=\bar n_0(z) \tilde C$, where $\bar n_0(z)=\int d^2\bm \rho d^3\bm p f_0(\bm r,\bm p)=\bar n_0(0) e^{-\omega_z^2 z^2/2}$ is the equilibrium 1D-density and $\tilde C$ no longer acts on the coordinate $z$. Taking $z'=\omega_z z$, we obtain
\be
\left[-\gamma+\omega_z(p_z\partial_{z'}-z'\partial_{p_z})\right]\widetilde\alpha =-{\cal L}_{\bar n_0(z')}[\widetilde\alpha ],
\label{Eq:4}
\ee
with
\be
{\cal L}_{\bar n_0(z')}=\bar n_0(z')\tilde C+\bm\Pi\cdot\partial_{\bm\rho}-\bm\rho\cdot\partial_{\bm\Pi}.
\ee
Note that ${\cal L}_{\bar n_0(z')}$ depends on the axial coordinate $z'$ only through the axial density. In particular, $z'$ is only a parameter of the operator since we neither integrate nor differentiate with respect to this coordinate.

Two properties of ${\cal L}_{\bar n}$ will be used below: (i) its kernel is spanned by the functions of $z'$ only \footnote{This result can be obtained by noting that if $\alpha$ belongs to the kernel of $C$ we have first $\langle \alpha|C[\alpha]\rangle=\langle \alpha|{\cal L}[\alpha]\rangle=0$ where the scalar product is defined as in Eq. (\ref{scalar}). Moreover we have $\langle \alpha|C[\alpha]\rangle=\int d^2\bm\rho d^3\bm p_1 d^3\bm p_2 f_{0,1}f_{0,2}v_{\rm rel}\sigma (\alpha_1-\alpha_2)^2/2$. So we see that $\alpha$ should not depend on the momentum and that $C[\alpha]=0$. Using these properties in the equation ${\cal L}[\alpha]=0$, we see that $\alpha$ is not a function of $\rho$ either.} and (ii) due to atom number conservation, we have for any $\widetilde\alpha$, $\int d^2\bm\rho d^3\bm p f_0 {\cal L}_{\bar n}[\widetilde\alpha ]=0$.
We look for solutions of Eq.~(\ref{Eq:4}) in the limit of a very elongated trap $\omega_z\rightarrow 0$. If we focus on the {\em slow} axial spin dynamics of the cloud studied experimentally in \cite{sommer2011universal}, we have also $\gamma\rightarrow 0$ and we can therefore expand $\gamma$ and $\widetilde\alpha $ as $\gamma=\sum_{n\ge 1}\gamma_n \omega_z^n$ and $\widetilde\alpha(\bm r,\bm p)=\sum_{n\ge 0}\omega_z^n a_n(\bm r,\bm p)$. Note that we are ultimately interested in the coefficient $\gamma_2$, since we take $\Gamma_{\rm SD}=\omega_z^2/\gamma$ as in \cite{sommer2011universal}.
Inserting these expansions in Eq.~(\ref{Eq:4}) we get to zero-th order ${\cal L}_{\bar n}[a_0]=0$ \footnote{Note that, strictly speaking, $\omega_z$ still appears in $C$ through the normalisation factor $\cal A$. We thus take the limit $\omega_z\rightarrow 0$ at constant peak-density to avoid this difficulty.}. According to property (i), $a_0$ is thus a function of $z'$ only. It is determined explicitly by the study of the next order terms of the expansion. For $n=1$ we obtain
\be
-\gamma_1 a_0 + p_z \partial_{z'}a_0=-{\cal L}_{\bar n_0(z')}[a_1].
\ee
Using property (ii), we see readily that $\gamma_1=0$, which then leads to the following relation:
\be
p_z \partial_{z'}a_0=-{\cal L}_{\bar n_0(z')}[a_1].
\label{Eq:4b}
\ee
Consider a uniform density $\bar n$ and assume for a moment that we know the solution $\chi_{\bar n} (\bm\rho,\bm p)$ of the integro-differential equation $p_z={\cal L}_{\bar n}[\chi_{\bar n}]$ (the properties of $\chi_{\bar n}$ will be discussed below). Since $\bar n_0 (z')$ is only a parameter, and by linearity of $\cal L$, the solution of Eq.~(\ref{Eq:4b}) is thus $a_1=-\chi_{\bar n_0(z')}\partial_{z'}a_0$.

Having expressed $a_1$ as a function of $a_0$, we close the set of equations by considering  the $n=2$ term of the expansion. It reads:
\be
\left(-\gamma_2 a_0 + (p_z\partial_{z'}-z'\partial_{p_z})a_1\right)=-{\cal L}_{\bar n_0(z')}[a_2].
\ee
Using the expression of $a_1$ as well as the property (ii), we obtain after integration by parts
\be
\gamma_2 \bar n_0(z')a_0(z')+\partial_{z'}\left(G(z')\partial_{z'}a_0(z')\right)=0,
\label{Eq:5}
\ee
where
\be
G(z')\equiv \int d^2\bm\rho d^3\bm p f_0 p_z \chi_{\bar n_0(z')}(\bm\rho,\bm p).
\ee

We now show that $G$ defined above can be identified with the spin conductance of an ideal gas of 1D density $\bar n$ in a cylindrical harmonic trap. Indeed, by definition, the conductance is obtained by solving Boltzmann's equation in a cylindrical trap in the presence of a spin pulling force $\bm F_s=s F_0\bm e_z$. Expanding Eq. (\ref{Eq:1}) to first order in perturbation, and taking as above $f_s=f_0 (1+s \alpha(\bm \rho,\bm p))$,  we see that $\alpha$ is solution of
\be
F_0p_z = {\cal L}_{\bar n}[\alpha].
\ee
 We recognize here the same equation as for the definition of $\chi$ and we then have $\alpha=F_0\chi_{\tilde n}$. Since the particle flux is given by $\Phi=\int d^3\bm pd^2\bm\rho f_0 \alpha p_z$, we see finally that, as claimed above, $G=\Phi/F_0=\int d^2\rho d^3\bm p f_0p_z\chi_{\tilde n}(\bm \rho,\bm p)$.
\section{Spin drag coefficient for the Maxwellian gas}

Consider the special case of the radially trapped Maxwellian gas for which $\sigma (p)=\Lambda/p$ where $\Lambda$ is some constant. This model is useful to interpolate between the weakly interacting ($\sigma=$const.) and the strongly interacting ($\sigma\propto1/p^2$) limits. Taking the ansatz $\alpha(\bm r,\bm p)=F_0 p_z H(\bm\rho,\bm\Pi)/2$, Boltzmann's equation for spin excitations  turns into
\be
\frac{1}{2}\left(\bm\Pi\cdot\partial_{\bm\rho}-\bm\rho\cdot\partial_{\bm\Pi}\right)H(\bm\rho,\bm\Pi)-1=-\frac{\Gamma_0}{2} e^{-\rho^2/2}H(\bm\rho,\bm\Pi),
\label{EqSOM1}
\ee
with $\Gamma_0=\Lambda n_0$ the damping rate of the spin excitations of a homogeneous gas of density $n_0$. Using the rotational invariance around the $z$ axis, the phase space density can be expressed using the  new variables
\begin{eqnarray*}
h&=&(\Pi^2+\rho^2)/2\\
u&=&(\Pi^2-\rho^2)/2\\
v&=&\bm\Pi\cdot\bm\rho.
\end{eqnarray*}
Let $u+iv=R e^{i\varphi}$. Eq. (\ref{EqSOM1}) then becomes
\be
\partial_{\varphi}H(h,R,\varphi)-1=-\frac{\Gamma_0 e^{-h/2}}{2} e^{R\cos\varphi/2}H(h,R,\varphi).
\label{EqSOM2}
\ee
Moreover, in these new variables, we have
\be
\int d^2\bm\Pi d^2\bm\rho \cdots=2\pi\int_{h=0}^\infty\int_{x=0}^1\int_{\varphi=0}^{2\pi}\frac{xdx\, hdh\,d\varphi}{\sqrt{1-x^2}} \cdots .
\ee
where $R=xh$ and the dots stand for any cylindrically symmetric function of $\bm\Pi$ and $\bm\rho$. In particular, the spin-conductance is given by
\be
G=\frac{n_0}{2}\int\frac{xdx\, hdh\,d\varphi}{\sqrt{1-x^2}} e^{-h}H(h,R=xh,\varphi),
\label{EqSOM4}
\ee
We now turn to the solution of Eq.~(\ref{EqSOM2}) where we focus on the $\varphi$-dependence, since it is the only variable appearing in the differential operator. Eq. (\ref{EqSOM2}) takes the general form
\be
H'(\varphi)+\mu A(\varphi) H(\varphi)=1,
\label{EqSOM3}
\ee
with $\mu=\Gamma_0 e^{-h/2}/2$ and $A=\exp(R\cos(\varphi)/2)$ is a $2\pi$-periodic function. Take
\be
K_{\mu}(\varphi)=\exp\left[-\mu\int_0^{\varphi}d\varphi' A(\varphi')\right],
\ee
the general solution of (\ref{EqSOM3}) is
\be
H(\varphi)=K_\mu(\varphi)\int_{\varphi_0}^\varphi\frac{d\varphi'}{K_\mu(\varphi')},
\ee
where $\varphi_0$ is an integration constant that can be determined by imposing the periodicity of $H$. Taking $H(0)=H(2\pi)$, we have finally
\be
H(\varphi)=K_\mu(\varphi)\left[\int_0^\varphi\frac{d\varphi'}{K_\mu(\varphi')}+\frac{K_{\mu}(2\pi)}{1-K_{\mu}(2\pi)}\int_0^{2\pi}\frac{d\varphi'}{K_\mu(\varphi')}\right].
\ee
Let's now discuss the behavior of the solutions in the collisionless and hydrodynamic limits.

\subsection{Collisionless limit $\mu\rightarrow 0$}
In this limit, $K_\mu\simeq 1-\mu\int_0^{\varphi}A(\varphi')d\varphi'$. The asymptotic behavior of $H$ is then dominated by the singularity due to the denominator $1-K_\mu(2\pi)$ that vanishes for $\mu=0$. To leading order, we see that $H$ does not depend on $\varphi$ and is given by
    \be
    H=\frac{1}{\mu\bar A},
    \ee
where
\be
\bar A=\frac{1}{2\pi}\int_0^{2\pi}A(\varphi')d\varphi'=I_0(R/2)
\ee
is the average value of $A$ ($I_0$ is the zeroth-order modified Bessel function of the first kind).

Using the actual values of $\mu$ and $A$ we have
\begin{eqnarray}
G&\sim&\frac{n_0}{\Gamma_0}\int_{h=0}^\infty\int_{x=0}^1\frac{hdh\,xdx}{\sqrt{1-x^2}} \frac{2\pi e^{-h/2}}{{\rm I}_0(xh/2)}\\
&\sim& \frac{15.87}{\Lambda},
\end{eqnarray}
Note in particular that $G$ does not depend on the density of the cloud.

Within this limit the velocity field in the collisionless regime is given by
\be
v(\rho)=
\frac{F_0}{2\pi\Gamma_0}\int \frac{e^{(\rho^2-\Pi^2)/4}\Pi d\Pi d\theta}{I_0\left(\frac{1}{2}\sqrt{\Pi^2\rho^2\cos^2\theta+(\Pi^2-\rho^2)^2/4}\right)}
\label{EqSOMMaxwell}
\ee

\subsection{Hydrodynamic limit $\mu\rightarrow\infty$}
In the limit $\Gamma_0\rightarrow\infty$ we may  neglect the transport term $H'$ in (\ref{EqSOM3}), yielding $H(\varphi)=1/\mu A(\varphi)\propto 1/\Gamma_0$.  Taking $\Gamma(\bm \rho)=\Gamma_0 e^{-\rho^2/2}$, the local spin damping rate, we recover the expected result $v(\rho )\propto 1/\Gamma (\bm\rho)$, that was obtained in the main text using local density arguments.

\section{A variational result for the collisionless spin conductance}

The spin conductance in a transverse trap is obtained by solving the linearized Boltzmann equation
\be
\left(\bm\Pi\cdot\partial_{\bm\rho}-\bm\rho\cdot\partial_{\bm\Pi}\right)\alpha-Fp_z=-C[\alpha],
\ee
where $C$ is the linearized collisional operator defined by
\be
C[\alpha](\bm\rho,\bm p_1)=\int d^3\bm p_2f_0(\bm\rho,\bm p_2)|\bm p_2-\bm p_1|\sigma \left[\alpha(\bm\rho,\bm p_1)-\alpha(\bm\rho,\bm p_2)\right].
\ee
We recall that $C$ is symmetric and positive for the scalar product
\be
\langle g_1|g_2\rangle=\int d^2\bm\rho d^3\bm p g_1(\bm\rho,\bm p)g_2(\bm\rho,\bm p)f_0(\bm\rho,\bm p), \label{scalar}
\ee
where $f_0(\bm\rho,\bm p)=n_0e^{-(p^2+\rho^2)/2}/\sqrt{2\pi}^3$ is the static phase-space density, and $n_0$  is the density at the center of the trap.

The spin current is defined by $\Phi=\int d^3\bm p d^2\bm\rho f_0 \alpha p_z=\langle \alpha|p_z\rangle$ and the spin conductance is then $G=\Phi/F$. Letting $\alpha(\bm\rho,\bm p)=F a(\bm\rho,\bm p)$ we have more simply $G=\langle p_z|a\rangle$.

We work in the collisionless limit $\sigma\rightarrow 0$ and we thus take $\sigma (p_{\rm rel})=\varepsilon \hat\sigma (p_{\rm rel})$ and $C=\varepsilon C_2$ where $\varepsilon$ is small. Following the results obtained for the Maxwellian gas, we expand $a$ as
\be
a(\bm\rho,\bm p)=\frac{a_0(\bm\rho,\bm p)}{\varepsilon}+a_1(\bm\rho,\bm p)+\varepsilon a_2(\bm\rho,\bm p)+\ldots
\ee
Inserting this expansion in Boltzmann's equation, we obtain to leading order
\be
\left(\bm\Pi\cdot\partial_{\bm\rho}-\bm\rho\cdot\partial_{\bm\Pi}\right)a_0=0.
\label{Eq3SOM}
\ee
This equation is solved readily by introducing the variables $(p_z,h,x,\varphi)$ defined in the study of the Maxwellian gas.

In these coordinates, Eq.~(\ref{Eq3SOM}) becomes simply $\partial_\varphi a_0=0$. The set ${\cal F}_0$ of solutions of Eq.~(\ref{Eq3SOM}) is thus composed of functions whose value does not depend on the angle $\varphi$. To get the actual expression of $a_0$ we need to go one step further in the expansion. At this order in $\varepsilon$, we have
\be
\partial_\varphi a_1-p_z=-C_2[a_0].
\ee
To get rid of $a_1$, we integrate over $\varphi$ and use the fact that the $a_n$ are periodic functions of $\varphi$. We then obtain the equation
\be
\bar C_2[a_0]=p_z,
\label{Eq2}
\ee
 with $\bar C_2[a_0]=\int d\varphi C_2[a_0]/2\pi$ and where $a_0$ is now the only unknown.

We define on ${\cal F}_0$ the new scalar product
\be
(a|b)=4\pi^2\int \frac{xdx\, hdh\,  dp_z}{\sqrt{1-x^2}} f_0 a(x,h,p_z)b(x,h,p_z),
\ee
which is equivalent to the old scalar product $\langle a|b\rangle$.
We then see readily that $(a|\bar C[b])=\langle a|C[b]\rangle$. Using the properties of $C$, we deduce that $\bar C_2$ is a symmetric, positive operator on ${\cal F}_0$. Eq. (\ref{Eq2}) then has the same structure as the ones used to calculate transport coefficients in homogeneous systems. We can then use the usual tricks to get a bound on the spin conductance \cite{smith1989transport}. We indeed write that for any real $\lambda$ and any function $b\in {\cal F}_0$, we have $( a_0+\lambda b|\bar C_2[a_0+\lambda b])\ge 0$, and using the fact that this second order polynomial in $\lambda$ is always positive, we obtain from the negativity of the discriminant that for any $b$,
\be
G\ge \frac{( p_z|b)^2}{( b|\bar C[b])}=\frac{\langle p_z|b\rangle^2}{\langle b|C[b]\rangle},
\ee
the bound being reached for $b=a_0$. We take as a variational ansatz $b=p_z$, since as discussed earlier, the collisionless regime is associated with a rather flat velocity profile. We then obtain
\be
G\ge \frac{\left(\int d^2\bm\rho e^{-\rho^2/2}\right)^2}{\int d^2\bm\rho e^{-\rho^2}}\frac{n_0}{\Gamma_0}=4\pi \frac{n_0}{\Gamma_0},
\ee
with $\Gamma_0$ the spin drag on the axis of the trap. The prefactor is $4\pi\simeq 12.56$, not far from the result 15.87 found analytically for the Maxwellian gas, and the bound is indeed satisfied.

An improved variational bound can be obtained by using the exact result Eq. (\ref{EqSOMMaxwell}) found for the Maxwellian gas to estimate the spin drag for constant-momentum or unitary-limited cross-sections. For the Maxwellian gas, this gives by definition the exact result, while for the constant cross-section and the unitary gases, we obtain respectively $\Gamma\ge 14.5/\Gamma_0$ and $\Gamma\ge 17/\Gamma_0$.

\section{Interpolation scheme for the spin conductance}

We know that for a power-law cross-section \footnote{For a real cross-section, $G$ should also depend on $k_{\rm th}a$ - although this dependence is weak.}, the spin conductance $G$ scales like $n_0(0)/\Gamma_0 f(1/\Gamma_0)$ where $f$ obeys the following asymptotic behaviors:

\begin{enumerate}
\item In the collisionless limit, $f$ converges to a constant value ($\simeq 15.87$ for the Maxwellian gas).
\item In the hydrodynamic limit, $f$ has a logarithmic singularity and scales like $2\pi \ln \Gamma_0$.
\end{enumerate}

\begin{figure}
\centerline{\includegraphics[width=\columnwidth]{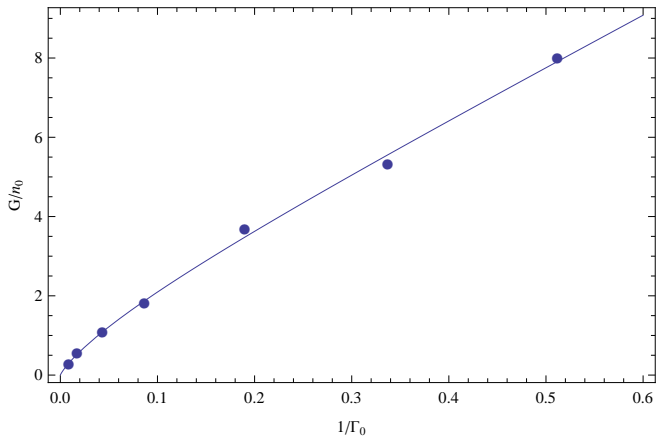}}
\caption{Spin conductance $G$ for a constant cross-section. Dots: Molecular Dynamics  Simulation. Solid line: Pad\'e's approximation. }
\label{Fig:1}
\end{figure}

To interpolate between these two limits we make use of the Bessel function $K_0$ which vanishes at $+\infty$ and diverges as $-\ln x$ at $x=0$. We thus approximate $f$ by
\be
f(x)=2\pi K_0(x)+15.87\frac{x+a}{x+b},
\ee
where $a$ and $b$ are determined by fitting the results of the molecular dynamics simulations (see Fig. \ref{Fig:1}). In the case of a constant cross-section, we obtain $a=0.11$ and $b=0.52$. We note that the largest relative error between the Pad\'e interpolation and the result of the molecular simulation is observed for the largest values of $\Gamma_0$ and amounts to $\simeq 6$~\% for $\Gamma_0\simeq 10$.


\bibliography{bibliographie}

\begin{thebibliography}{24}
\expandafter\ifx\csname natexlab\endcsname\relax\def\natexlab#1{#1}\fi
\expandafter\ifx\csname bibnamefont\endcsname\relax
  \def\bibnamefont#1{#1}\fi
\expandafter\ifx\csname bibfnamefont\endcsname\relax
  \def\bibfnamefont#1{#1}\fi
\expandafter\ifx\csname citenamefont\endcsname\relax
  \def\citenamefont#1{#1}\fi
\expandafter\ifx\csname url\endcsname\relax
  \def\url#1{\texttt{#1}}\fi
\expandafter\ifx\csname urlprefix\endcsname\relax\def\urlprefix{URL }\fi
\providecommand{\bibinfo}[2]{#2}
\providecommand{\eprint}[2][]{\url{#2}}

\bibitem[{\citenamefont{Zwerger}(2012)}]{zwerger2012BCSBEC}
\bibinfo{editor}{\bibfnamefont{W.}~\bibnamefont{Zwerger}}, ed.,
  \emph{\bibinfo{title}{The BCS-BEC Crossover and the Unitary Fermi Gas}}, vol.
  \bibinfo{volume}{836} of \emph{\bibinfo{series}{Lecture Notes in Physics}}
  (\bibinfo{publisher}{Springer}, \bibinfo{address}{Berlin},
  \bibinfo{year}{2012}).

\bibitem[{\citenamefont{Brantut et~al.}(2012)\citenamefont{Brantut, Meineke,
  Stadler, Krinner, and Esslinger}}]{brantut2012conduction}
\bibinfo{author}{\bibfnamefont{J.-P.} \bibnamefont{Brantut}},
  \bibinfo{author}{\bibfnamefont{J.}~\bibnamefont{Meineke}},
  \bibinfo{author}{\bibfnamefont{D.}~\bibnamefont{Stadler}},
  \bibinfo{author}{\bibfnamefont{S.}~\bibnamefont{Krinner}}, \bibnamefont{and}
  \bibinfo{author}{\bibfnamefont{T.}~\bibnamefont{Esslinger}},
  \bibinfo{journal}{Science} \textbf{\bibinfo{volume}{337}},
  \bibinfo{pages}{1069} (\bibinfo{year}{2012}).

\bibitem[{\citenamefont{Cao et~al.}(2011)\citenamefont{Cao, Elliott, Joseph,
  Wu, Petricka, Sch{\"a}fer, and Thomas}}]{cao2010observation}
\bibinfo{author}{\bibfnamefont{C.}~\bibnamefont{Cao}},
  \bibinfo{author}{\bibfnamefont{E.}~\bibnamefont{Elliott}},
  \bibinfo{author}{\bibfnamefont{J.}~\bibnamefont{Joseph}},
  \bibinfo{author}{\bibfnamefont{H.}~\bibnamefont{Wu}},
  \bibinfo{author}{\bibfnamefont{J.}~\bibnamefont{Petricka}},
  \bibinfo{author}{\bibfnamefont{T.}~\bibnamefont{Sch{\"a}fer}},
  \bibnamefont{and} \bibinfo{author}{\bibfnamefont{J.~E.}
  \bibnamefont{Thomas}}, \bibinfo{journal}{Science}
  \textbf{\bibinfo{volume}{331}}, \bibinfo{pages}{58} (\bibinfo{year}{2011}),
  ISSN \bibinfo{issn}{0036-8075}.

\bibitem[{\citenamefont{Kovtun et~al.}(2005)\citenamefont{Kovtun, Son, and
  Starinets}}]{kovtun2005viscosity}
\bibinfo{author}{\bibfnamefont{P.}~\bibnamefont{Kovtun}},
  \bibinfo{author}{\bibfnamefont{D.~T.} \bibnamefont{Son}}, \bibnamefont{and}
  \bibinfo{author}{\bibfnamefont{A.}~\bibnamefont{Starinets}},
  \bibinfo{journal}{Phys. Rev. Lett.} \textbf{\bibinfo{volume}{94}},
  \bibinfo{pages}{111601} (\bibinfo{year}{2005}).

\bibitem[{\citenamefont{Liao et~al.}(2011)\citenamefont{Liao, Revelle,
  Paprotta, Rittner, Li, Partridge, and Hulet}}]{liao2011metastability}
\bibinfo{author}{\bibfnamefont{Y.}~\bibnamefont{Liao}},
  \bibinfo{author}{\bibfnamefont{M.}~\bibnamefont{Revelle}},
  \bibinfo{author}{\bibfnamefont{T.}~\bibnamefont{Paprotta}},
  \bibinfo{author}{\bibfnamefont{A.}~\bibnamefont{Rittner}},
  \bibinfo{author}{\bibfnamefont{W.}~\bibnamefont{Li}},
  \bibinfo{author}{\bibfnamefont{G.}~\bibnamefont{Partridge}},
  \bibnamefont{and} \bibinfo{author}{\bibfnamefont{R.}~\bibnamefont{Hulet}},
  \bibinfo{journal}{Phys. Rev. Lett.} \textbf{\bibinfo{volume}{107}},
  \bibinfo{pages}{145305} (\bibinfo{year}{2011}).

\bibitem[{\citenamefont{Bruun}(2012)}]{bruun2012shear}
\bibinfo{author}{\bibfnamefont{G.}~\bibnamefont{Bruun}},
  \bibinfo{journal}{Phys. Rev. A} \textbf{\bibinfo{volume}{85}},
  \bibinfo{pages}{013636} (\bibinfo{year}{2012}).

\bibitem[{\citenamefont{Enss and Haussmann}(2012)}]{enss2012quantum}
\bibinfo{author}{\bibfnamefont{T.}~\bibnamefont{Enss}} \bibnamefont{and}
  \bibinfo{author}{\bibfnamefont{R.}~\bibnamefont{Haussmann}},
  \bibinfo{journal}{Phys. Rev. Lett.} \textbf{\bibinfo{volume}{109}},
  \bibinfo{pages}{195303} (\bibinfo{year}{2012}).

\bibitem[{\citenamefont{Wong et~al.}(2012)\citenamefont{Wong, Stoof, and
  Duine}}]{wong2012spin}
\bibinfo{author}{\bibfnamefont{C.}~\bibnamefont{Wong}},
  \bibinfo{author}{\bibfnamefont{H.}~\bibnamefont{Stoof}}, \bibnamefont{and}
  \bibinfo{author}{\bibfnamefont{R.}~\bibnamefont{Duine}},
  \bibinfo{journal}{Phys. Rev. A} \textbf{\bibinfo{volume}{85}},
  \bibinfo{pages}{063613} (\bibinfo{year}{2012}).

\bibitem[{\citenamefont{Heiselberg}(2012)}]{heiselberg2012inhomogeneous}
\bibinfo{author}{\bibfnamefont{H.}~\bibnamefont{Heiselberg}},
  \bibinfo{journal}{Phys. Rev. Lett.} \textbf{\bibinfo{volume}{108}},
  \bibinfo{pages}{245303} (\bibinfo{year}{2012}).

\bibitem[{\citenamefont{Kittinaradorn et~al.}(2012)\citenamefont{Kittinaradorn,
  Duine, and Stoof}}]{kittinaradorn2012critical}
\bibinfo{author}{\bibfnamefont{R.}~\bibnamefont{Kittinaradorn}},
  \bibinfo{author}{\bibfnamefont{R.}~\bibnamefont{Duine}}, \bibnamefont{and}
  \bibinfo{author}{\bibfnamefont{H.}~\bibnamefont{Stoof}},
  \bibinfo{journal}{New Journal of Physics} \textbf{\bibinfo{volume}{14}},
  \bibinfo{pages}{055007} (\bibinfo{year}{2012}).

\bibitem[{\citenamefont{Kim and Huse}(2012)}]{kim2012heat}
\bibinfo{author}{\bibfnamefont{H.}~\bibnamefont{Kim}} \bibnamefont{and}
  \bibinfo{author}{\bibfnamefont{D.~A.} \bibnamefont{Huse}},
  \bibinfo{journal}{Phys. Rev. A} \textbf{\bibinfo{volume}{86}},
  \bibinfo{pages}{053607} (\bibinfo{year}{2012}).

\bibitem[{\citenamefont{Musaelian and Meyerovich}(1992)}]{Meyerovich}
\bibinfo{author}{\bibfnamefont{K.}~\bibnamefont{Musaelian}} \bibnamefont{and}
  \bibinfo{author}{\bibfnamefont{A.}~\bibnamefont{Meyerovich}},
  \bibinfo{journal}{Journal of Low Temperature Physics}
  \textbf{\bibinfo{volume}{89}}, \bibinfo{pages}{535} (\bibinfo{year}{1992}),
  ISSN \bibinfo{issn}{0022-2291},
  \urlprefix\url{http://dx.doi.org/10.1007/BF00694081}.

\bibitem[{\citenamefont{Mineev}(2004)}]{Mineev}
\bibinfo{author}{\bibfnamefont{V.~P.} \bibnamefont{Mineev}},
  \bibinfo{journal}{Phys. Rev. B} \textbf{\bibinfo{volume}{69}},
  \bibinfo{pages}{144429} (\bibinfo{year}{2004}),
  \urlprefix\url{http://link.aps.org/doi/10.1103/PhysRevB.69.144429}.

\bibitem[{\citenamefont{Wolf et~al.}(2001)\citenamefont{Wolf, Awschalom,
  Buhrman, Daughton, Von~Molnar, Roukes, Chtchelkanova, and
  Treger}}]{wolf2001spintronics}
\bibinfo{author}{\bibfnamefont{S.}~\bibnamefont{Wolf}},
  \bibinfo{author}{\bibfnamefont{D.}~\bibnamefont{Awschalom}},
  \bibinfo{author}{\bibfnamefont{R.}~\bibnamefont{Buhrman}},
  \bibinfo{author}{\bibfnamefont{J.}~\bibnamefont{Daughton}},
  \bibinfo{author}{\bibfnamefont{S.}~\bibnamefont{Von~Molnar}},
  \bibinfo{author}{\bibfnamefont{M.}~\bibnamefont{Roukes}},
  \bibinfo{author}{\bibfnamefont{A.~Y.} \bibnamefont{Chtchelkanova}},
  \bibnamefont{and} \bibinfo{author}{\bibfnamefont{D.}~\bibnamefont{Treger}},
  \bibinfo{journal}{Science} \textbf{\bibinfo{volume}{294}},
  \bibinfo{pages}{1488} (\bibinfo{year}{2001}).

\bibitem[{\citenamefont{Sommer et~al.}(2011{\natexlab{a}})\citenamefont{Sommer,
  Ku, Roati, and Zwierlein}}]{sommer2011universal}
\bibinfo{author}{\bibfnamefont{A.}~\bibnamefont{Sommer}},
  \bibinfo{author}{\bibfnamefont{M.}~\bibnamefont{Ku}},
  \bibinfo{author}{\bibfnamefont{G.}~\bibnamefont{Roati}}, \bibnamefont{and}
  \bibinfo{author}{\bibfnamefont{M.~W.} \bibnamefont{Zwierlein}},
  \bibinfo{journal}{Nature} \textbf{\bibinfo{volume}{472}},
  \bibinfo{pages}{201} (\bibinfo{year}{2011}{\natexlab{a}}).

\bibitem[{\citenamefont{Sommer et~al.}(2011{\natexlab{b}})\citenamefont{Sommer,
  Ku, and Zwierlein}}]{summer2011spin}
\bibinfo{author}{\bibfnamefont{A.}~\bibnamefont{Sommer}},
  \bibinfo{author}{\bibfnamefont{M.}~\bibnamefont{Ku}}, \bibnamefont{and}
  \bibinfo{author}{\bibfnamefont{M.~W.} \bibnamefont{Zwierlein}},
  \bibinfo{journal}{New Journal of Physics} \textbf{\bibinfo{volume}{13}},
  \bibinfo{pages}{055009} (\bibinfo{year}{2011}{\natexlab{b}}).

\bibitem[{\citenamefont{Bruun and Pethick}(2011)}]{bruun2011spinb}
\bibinfo{author}{\bibfnamefont{G.}~\bibnamefont{Bruun}} \bibnamefont{and}
  \bibinfo{author}{\bibfnamefont{C.}~\bibnamefont{Pethick}},
  \bibinfo{journal}{Phys. Rev. Lett.} \textbf{\bibinfo{volume}{107}},
  \bibinfo{pages}{255302} (\bibinfo{year}{2011}).

\bibitem[{\citenamefont{Vichi and Stringari}(1999)}]{vichi1999collective}
\bibinfo{author}{\bibfnamefont{L.}~\bibnamefont{Vichi}} \bibnamefont{and}
  \bibinfo{author}{\bibfnamefont{S.}~\bibnamefont{Stringari}},
  \bibinfo{journal}{Phys. Rev. A} \textbf{\bibinfo{volume}{60}},
  \bibinfo{pages}{4734} (\bibinfo{year}{1999}).

\bibitem[{\citenamefont{Chiacchiera et~al.}(2010)\citenamefont{Chiacchiera,
  Macri, and Trombettoni}}]{chiacchiera2010dipole}
\bibinfo{author}{\bibfnamefont{S.}~\bibnamefont{Chiacchiera}},
  \bibinfo{author}{\bibfnamefont{T.}~\bibnamefont{Macri}}, \bibnamefont{and}
  \bibinfo{author}{\bibfnamefont{A.}~\bibnamefont{Trombettoni}},
  \bibinfo{journal}{Phys. Rev. A} \textbf{\bibinfo{volume}{81}},
  \bibinfo{pages}{033624} (\bibinfo{year}{2010}).

\bibitem[{\citenamefont{Bruun}(2011)}]{bruun2011spin}
\bibinfo{author}{\bibfnamefont{G.~M.} \bibnamefont{Bruun}},
  \bibinfo{journal}{New Journal of Physics} \textbf{\bibinfo{volume}{13}},
  \bibinfo{pages}{035005} (\bibinfo{year}{2011}).

\bibitem[{\citenamefont{Goulko et~al.}(2011)\citenamefont{Goulko, Chevy, and
  Lobo}}]{goulko2011collision}
\bibinfo{author}{\bibfnamefont{O.}~\bibnamefont{Goulko}},
  \bibinfo{author}{\bibfnamefont{F.}~\bibnamefont{Chevy}}, \bibnamefont{and}
  \bibinfo{author}{\bibfnamefont{C.}~\bibnamefont{Lobo}},
  \bibinfo{journal}{Phys. Rev. A} \textbf{\bibinfo{volume}{84}},
  \bibinfo{pages}{051605} (\bibinfo{year}{2011}).

\bibitem[{\citenamefont{Goulko et~al.}(2012)\citenamefont{Goulko, Chevy, and
  Lobo}}]{goulko2012boltzmann}
\bibinfo{author}{\bibfnamefont{O.}~\bibnamefont{Goulko}},
  \bibinfo{author}{\bibfnamefont{F.}~\bibnamefont{Chevy}}, \bibnamefont{and}
  \bibinfo{author}{\bibfnamefont{C.}~\bibnamefont{Lobo}}, \bibinfo{journal}{New
  Journal of Physics} \textbf{\bibinfo{volume}{14}}, \bibinfo{pages}{073036}
  (\bibinfo{year}{2012}).

\bibitem[{\citenamefont{Press et~al.}(2007)\citenamefont{Press, Teukolsky,
  Vetterling, and Flannery}}]{press2007numerical}
\bibinfo{author}{\bibfnamefont{W.~H.} \bibnamefont{Press}},
  \bibinfo{author}{\bibfnamefont{S.~A.} \bibnamefont{Teukolsky}},
  \bibinfo{author}{\bibfnamefont{W.~T.} \bibnamefont{Vetterling}},
  \bibnamefont{and} \bibinfo{author}{\bibfnamefont{B.~P.}
  \bibnamefont{Flannery}}, \emph{\bibinfo{title}{Numerical recipes 3rd edition:
  The art of scientific computing}} (\bibinfo{publisher}{Cambridge university
  press}, \bibinfo{year}{2007}).

\bibitem[{\citenamefont{Smith and Jensen}(1989)}]{smith1989transport}
\bibinfo{author}{\bibfnamefont{H.}~\bibnamefont{Smith}} \bibnamefont{and}
  \bibinfo{author}{\bibfnamefont{H.~H.} \bibnamefont{Jensen}},
  \emph{\bibinfo{title}{Transport phenomena}} (\bibinfo{publisher}{Oxford
  University Press, USA}, \bibinfo{year}{1989}).

\end{thebibliography}

\end{document}